%% file: main.tex
\documentclass[twocolumn, superscriptaddress]{revtex4}
\input{preamble}
\usepackage{tikz}
\usepackage{verbatim}

\begin{document}


\title{Probabilistic error cancellation for dynamic quantum circuits}

\author{Riddhi S. Gupta}
\affiliation{IBM Quantum, T. J. Watson Research Center, Yorktown Heights, New York 10598, USA}

\author{Ewout van den Berg}
\affiliation{IBM Quantum, T. J. Watson Research Center, Yorktown Heights, New York 10598, USA}

\author{Maika Takita}
\affiliation{IBM Quantum, T. J. Watson Research Center, Yorktown Heights, New York 10598, USA}

\author{Diego Riste}
\affiliation{IBM Quantum, T. J. Watson Research Center, Yorktown Heights, New York 10598, USA}

\author{Kristan Temme}
\affiliation{IBM Quantum, T. J. Watson Research Center, Yorktown Heights, New York 10598, USA}

\author{Abhinav Kandala}
\affiliation{IBM Quantum, T. J. Watson Research Center, Yorktown Heights, New York 10598, USA}

\begin{abstract}

Probabilistic error cancellation (PEC) is a technique that generates error-mitigated estimates of expectation values from ensembles of quantum circuits. In this work we extend the application of PEC from unitary-only circuits to dynamic circuits with measurement-based operations, such as mid-circuit measurements and classically-controlled (feedforward) Clifford operations. Our approach extends the sparse Pauli-Lindblad noise model to measurement-based operations while accounting for non-local measurement crosstalk in superconducting processors. Our mitigation and monitoring experiments provide a holistic view for the performance of the protocols developed in this work. These capabilities will be a crucial tool in the exploration of near-term dynamic circuit applications.
\end{abstract}
\maketitle


Extracting the most out of near-term quantum processors requires careful consideration of noise and hardware constraints. In recent years, the use of mid-circuit measurement and conditional quantum operations (`feedforward') has been shown to provide a distinct benefit over unitary-only circuits in enabling efficient computations in experiment, encompassing a variety of applications in topological quantum computing, quantum simulations and machine learning \cite{sivak_real-time_2023,gupta_encoding_2023,foss-feig_experimental_2023,cong_quantum_2019,baumer2023efficient,chen2023realizing}. However, mid-circuit measurements and real-time control flow often introduce new sources of noise that are not yet well understood. Indeed in the absence of practicable quantum error correction, quantum algorithms and applications must rely on error characterization \cite{beale_randomized_2023,govia_randomized_2022,Blumoff2016Implementing,Rudinger2022Characterizing,Pereira2022Complete,Pereira_2023} and mitigation techniques \cite{Temme_2017,berg_probabilistic_2022,berg_model-free_2022,tepaske_compressed_2023,botelho_error_2022,kim2023evidence} to address noise in order to fully realize the benefits of dynamic quantum circuits.

Probabilistic error cancellation (PEC) is a technique to learn and mitigate intrinsic noise in quantum channels. As introduced in  Refs.~\cite{Temme_2017,berg_probabilistic_2022} 
the protocol can compute unbiased estimators of observable expectation values by sampling noisy circuits from a probability distribution that is related to noise in the circuit. In conventional implementations of PEC ~\cite{song2019quantum,zhang2020error,berg_probabilistic_2022}, the target circuit operations have been unitary in the absence of noise, and measurements are pushed to the end of the computation. 
In this work, we extend this technique to quantum channels  where the target channel consists of mid-circuit measurements as well as Pauli operations conditioned on these measurements (`feedforward'). We enforce that these mid-circuit measurements are projective in the desired basis and our procedure ensures that these noise channels can be factorized into an ideal channel and intrinsic noise suitably twirled to yield Pauli channels. Our method characterizes the Pauli channel by solving a full-rank inverse problem to extract unique model coefficients from measured non-zero Pauli fidelities. Subsequently, the mitigation procedure is a direct extension to what was presented in Ref.~\cite{berg_probabilistic_2022}.

In what follows we outline the measurement-based learning protocol and present experimental results to confirm its efficacy as a new component of the PEC framework. Experimentally we find that, unlike two-qubit gate noise, measurement-induced noise is not limited to nearest-neighbor interactions. By using the appropriate readout topology to define a sparse Pauli Lindblad noise model, we mitigate noise for a joint channel consisting of mid-circuit measurements (on ancilla qubits) and unmeasured (data) qubits.  Under a unified PEC protocol, we further experimentally demonstrate error mitigation for a seven-qubit Clifford circuit consisting of unitary operations and mid-circuit measurements. Moving beyond mid-circuit measurements, we learn noise for a single feedforward operation and perform error mitigation on dynamic circuits. 

\section{Background \label{sec:background}}

The PEC framework ~\cite{Temme_2017} estimates noise-free observable expectation values by linearly combining noisy circuits sampled from an appropriately constructed probability distribution that, on average, cancels the effect of intrinsic noise. The protocol in \cite{berg_probabilistic_2022} assumes that intrinsic noise can be reshaped to Pauli noise via twirling \cite{bennett_purification_1996,knill_fault-tolerant_2004,kern_quantum_2005,geller_efficient_2013,wallman_noise_2016}. We use a Pauli-Lindblad parameterization of the Pauli channel where the noise model is expressed as a product of individual, commuting Pauli channels. Each channel is associated with a set of Pauli terms, $\mathcal{K}$ indexed by $l$, with associated  sample coefficients $w_l$, and model coefficients $\lambda_l \geq 0$. Following  Ref. \cite{berg_probabilistic_2022} we write a Pauli-Lindblad noise model as,
\begin{align}
    \Lambda(\rho) &= \mathop{\bigcirc}_{l\in \mathcal{K}} \left( w_l \cdot + (1-w_l)P_l \cdot P_l^\dagger \right)\rho, \label{eqn:def:pecnoisemodel}\\
     w_l &:= \frac{1 + \exp(-2\lambda_l)}{2}, \label{eqn:def:wk}
\end{align} where $l \in \mathcal{K}$ indexes all the Pauli terms in the noise model, $\mathop{\bigcirc}$ denotes a composition of maps, and $\cdot$ denotes a placeholder such that $T(\cdot)\rho = T(\rho)$. Subsequently, the Pauli fidelity $f_q$ measured in the Pauli basis specified by the set $\mathcal{F}$ and indexed by $q \in \mathcal{F}$ is, 
\begin{align}
    f_q = \exp \left(-2 \sum_{\langle q,l\rangle_{sp}=1} \lambda_l \right)
\end{align} where $\langle q,l\rangle_{sp}$ is unity if Pauli terms anticommute, $\{P_q, P_l\}=0$ and zero otherwise. Techniques for estimating Pauli fidelities of a twirled channel are well established \cite{kimmel_robust_2014,erhard_characterizing_2019,flammia_efficient_2020,helsen_new_2019} where the central idea is to measure fidelities after repeated applications of the channel in some Pauli basis. An exponential decay of the measured fidelity after $k$ repetitions of the channel is fit to the form $A f_{q}^k$ to estimate state preparation and measurement error, $A$ and learned fidelity, $f_{q}$.

The model coefficients, $\vec{\lambda}$, are related to the Pauli fidelities, $\vec{f}$ by the system of equations, 
\begin{align}
    -\log(\vec{f})/2 = M\vec{\lambda}, 
\end{align}
where matrix elements, $M_{q,l} = \langle q,l\rangle_{sp}$ store commutation relations for $P_q, P_l$ for all $q \in \mathcal{F}, l \in \mathcal{K}$.
We fit the model by extracting the model coefficients $\vec{\lambda} \geq 0$ from the list of learned fidelities by solving a non-negative least-squared minimization corresponding to this linear system.

\begin{figure}
    \centering
    \includegraphics[scale=0.9]{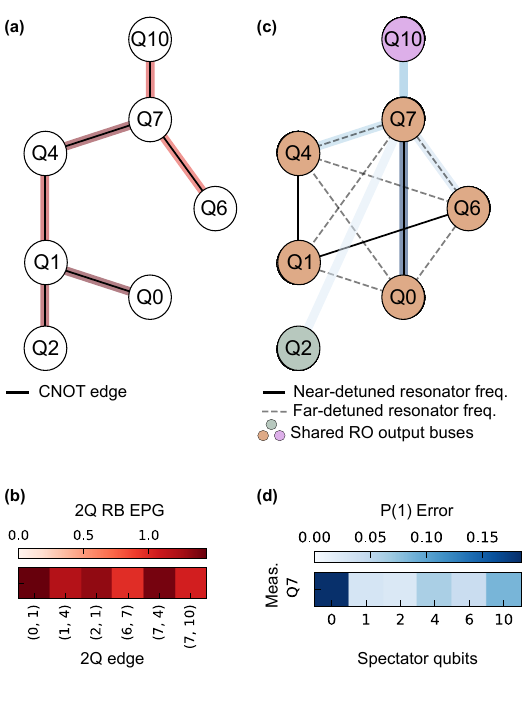}
    \caption{Sparse noise-model topologies for gates ((a),(b)) and measurements  ((c),(d)) on $\mathrm{ibm\_sapporo}$ (see \cref{app:devicedata}). (a) CNOT gate on physically connected edges (black solid) characterized by 2Q randomized benchmarking  with error per gate (red shaded) plotted in (b). These gate interactions are typically nearest-neighbor and provide a way to impose sparsity on noise models for unitary PEC. (c) Frequency multiplexed readout buses on the device (shaded circles). Within a single output bus (orange circles) qubit pairs whose measurement resonator frequencies are within $<65$MHz in resonator-frequency space (solid lines) are compared to qubit-pairs with far-detuned measurement resonators (dashed lines). The probability of the data (spectator) qubit in $|1\rangle$ state, $P(1)$, is measured via a pulse-sequence $X_{\pi/2}$-$\tau/2$-$X_{\pi}$-$\tau/2$-$Y_{\pi/2}$ for a fixed delay $\tau$ while a concurrent, high-amplitude measurement pulse is applied on the ancilla qubit (e.g. Q7) during the first delay period only. Shaded edges represent absolute error between measured and ideal $P(1)$ for qubit pairs in (d). (d) As an illustration, deviations from ideal $P(1)=0.5$ (blue shaded) are plotted for all spectators when Q7 is measured with fixed amplitude. Dominant pairwise measurement-induced crosstalk is anticipated for nearest-neighbours in resonator frequency space for the same output bus (e.g. Q0,Q7). Of all edges in (a), crosstalk in (c) is only observed for Q10,Q7 for a high measurement amplitude on Q7.}
    \label{fig:fig1}
\end{figure}

\begin{figure*}
    \centering
    \includegraphics[width=\textwidth]{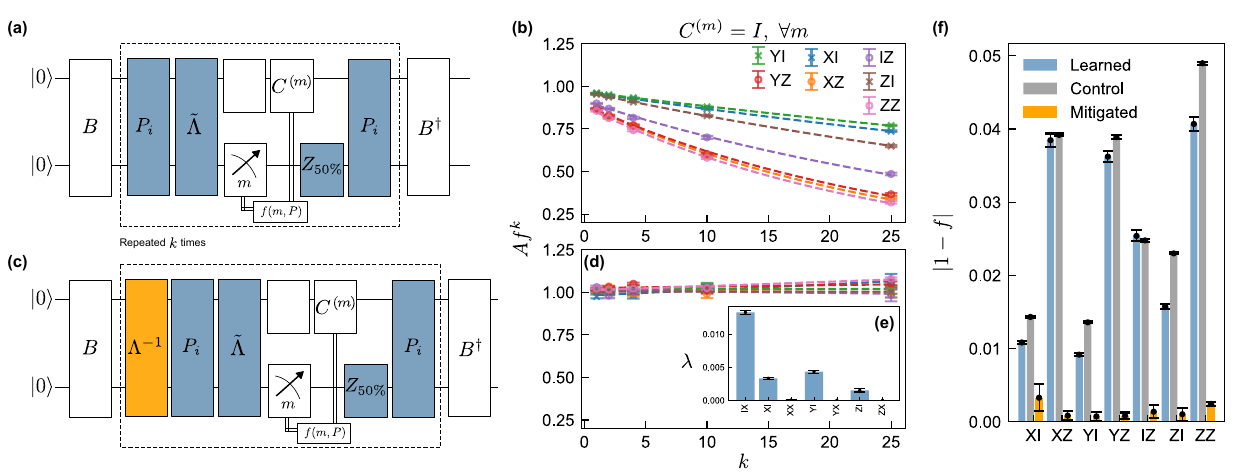}
    \caption{Learning and mitigation validation for two qubits. (a) The learning protocol for a measurement-based channel with intrinsic noise, $\tilde{\Lambda}$, mid-circuit measurement outcome $m$, and a feedforward operation $C^{(m)}$ implemented by controller $f(m,P)$. We add a $Z$-gate with 50\% probability ($Z_{50\%}$) to completely dephase post-measurement states and additionally perform Pauli-twirling ($P_i$). The twirled channel is repeated $k$ times and measured in a Pauli basis, $B$, yielding a measurement of eigenvalues $f_q, q \in \mathcal{F}$ of a diagonalized Pauli transfer matrix. (b) Decay in $Af_q^k$ versus depth $k$ using 256 circuit samples with 128 shots/circuit for all $B$ with non-zero $f_q, A$. Experimental data (fitted predictions) in markers (dashed lines) using Q0,7 on $\mathrm{ibm\_sapporo}$ where Q7 is measured without feedforward. Inset (e) depicts model coefficients $\lambda_{l},  l \in \mathcal{K}$ from fitted fidelities $f_q$ inverted using non-negative least squares. (c)  A mitigation validation experiment, where learning circuits in (a) are modified by inserting a mitigating layer of Paulis sampled from the inverse noise distribution (orange). (d) Decay in $Af_q^k$ versus depth $k$, where the number of random circuit samples are increased exponentially $\propto\gamma^{2k}$ for $k$ repeated applications. In contrast to (b), fidelity decay is suppressed by insertion of a mitigation layer for all Pauli bases. (f) Deviations of fitted $f_q$ from unity versus Pauli basis plotted for learning circuits (blue); mitigation-validation circuits (orange), and learning circuits interleaved during mitigation-validation circuits (gray). Error bars represent estimates of standard deviation from bootstrapping. Batches of learning and mitigation validation experiments are repeated every three hours for 13 trials. }
    \label{fig:fig2}
\end{figure*}

Based on these learned model coefficients, PEC is carried out by inserting random Paulis sampled from an appropriately constructed `inverse' distribution. The inverse, $\Lambda^{-1}$, of the overall channel reduces to the product of the (commuting) individual inverse channels, 
\begin{align}
     \Lambda^{-1}(\rho) &:= \gamma \mathop{\bigcirc}_{l\in \mathcal{K}} \left( w_l \cdot - (1-w_l)P_l \cdot P_l^\dagger \right)\rho, \label{eqn:def:inverse} \\
     \gamma  &:= \exp\left(\sum_{l \in \mathcal{K}} 2 \lambda_k\right),
\end{align} where $\gamma$ is a product of individual normalizing factors for each inverse channel. The sampling overhead required to achieve a unbiased estimate of the noiseless expectation value to some target uncertainty is $\gamma^2$ and $\gamma \geq 1$ equals unity only in the noiseless case.

As outlined in Ref.~\cite{berg_probabilistic_2022}, the net effect of inverting the average evolution of the channel under noise is achieved by inserting Pauli terms sampled from a re-normalized inverse distribution and appropriately re-scaling counts data in classical post-processing. 

\section{PEC for measurement-based circuits (MPEC) \label{sec:mpec}}

In this section, we extend the learning procedure such that one can learn joint linear channels containing unitary operations, mid-circuit measurements and feedforward. Of particular interest in measurement-based PEC is an appropriate choice of a sparse noise model via the set of noise generators $\mathcal{K}$. Unlike two-qubit gate noise in unitary PEC, measurement-induced noise on data qubits is not naturally limited to qubit-qubit nearest-neighbor  interactions. Details about the device readout configuration combined with calibration experiments are both used to reveal the relevant topology for defining $\mathcal{K}$ prior to an experiment, illustratively discussed in \cref{fig:fig1}.

We define a measurement-based ideal channel on both data (spectator) and ancilla (measured) qubits as,
\begin{align}
    \xi(\rho) := \sum_{m} C^{(m)} \Pi_m \rho \Pi_m  {C^\dagger}^{(m)}, 
\end{align} where $\Pi_m$ are projectors measuring only a subset of (ancilla) qubits satisfying  $\Pi_m \Pi_{m'} =\delta_{m, m'}$, $\sum_m \Pi_m = 1$ and $ \Pi_m=\Pi_m^\dagger$, while $C^{(m)}$ is a Pauli feedforward operation associated with classical ancilla outcomes $m$. In \cref{app:tmbc:cliffordffwd} we further argue any classically controlled Clifford circuit can be decomposed into unitary and non-unitary PEC layers where feedforward operations are at most single-qubit Clifford feedforward operations, for which MPEC continues to hold. The overhead of learning the noise for each conditional post-measurement branch \cite{beale_randomized_2023,singh2023experimental} can therefore be circumvented at the expense of inserting additional CNOT gates. We enforce factorizability of this ideal channel with noise,
\begin{align}
   \tilde{\xi}(\rho):=( \xi \circ \tilde{\Lambda})(\rho),
\end{align} by Pauli $Z$ twirling (denoted by $Z_{50\%}$ in \cref{fig:fig2}(a)), which applies a completely dephasing channel to the measured qubits. Namely, we apply a Pauli $Z$ with $50\%$ probability immediately after the measurement. This procedure acts identically to an ideal measurement in the $Z$ basis and enforces that learning noise generators corresponding to phase errors on the post-measurement state \cite{Govia2016entanglement} is not relevant to our procedure.

The resulting channel is subsequently twirled over the full Pauli group such that the reshaped intrinsic noise is represented by a diagonalized Pauli transfer matrix. \cref{app:tmbc:twirlingmsmts} shows that Pauli twirling elements can be pushed through measurement operations \cite{beale_randomized_2023}. In particular the  Pauli twirl applied on either side of the measured qubit is identical, and we additionally flip the classical outcome on each bit if the twirl was an $X$ or a $Y$. Ignoring any classical state discrimination error during readout, one obtains from \cref{eqn:twirledchannel} that
\begin{align}
    \mathrm{Tw}[\tilde{\xi}] 
    &=\Sigma_{m \in \mathcal{M}}  C^{(m)} \Pi_{(m)} \mathbb{E}\left[ P_i^\dagger \tilde{\Lambda}(P_i \cdot P_i^\dagger)  P_i \right]_i \Pi_{(m)}^\dagger  C^{(m)}, 
\end{align} where $\mathrm{Tw}[\cdot]$ represents twirling operation applied to an input channel; $P_i$ represent elements sampled uniformly and randomly from the Pauli group, and $\mathbb{E}$ denotes an averaging over these elements. In the above, we note that while the intrinsic noise is diagonalized, the combined Pauli transfer matrix of the channel and the noise is not diagonal. This issue can be addressed by noting that single qubit gates are assumed to be noiseless. Pauli feedforward operations can thus be replaced during learning by delays whose duration matches that of a single-qubit gate. Even though the feedforward operations are trivially identity, these operations trigger real-time control flow processing on hardware, for example, introducing mid-circuit idling times of order $700$ns on $\mathrm{ibm\_peekskill}$.  With $C^{(m)}  \equiv I$ for all $m$, the full channel is diagonal and takes the form,
\begin{align}
    \Sigma_{m \in \mathcal{M}} \Pi_{(m)} \mathbb{E} \left[  P_i^\dagger \tilde{\Lambda}(P_i \cdot P_i^\dagger) P_i  \right]_i \Pi_{(m)}^\dagger.
\end{align}

We summarize the learning protocol in \cref{fig:fig2}. The learning protocol for measuring fidelities in some Pauli basis $B$ specified by the set $\mathcal{F}$ is shown in \cref{fig:fig2}(a). An exponential decay of the measured fidelity is observed after $k$ repetitions of the channel. These measurements are fitted with a decay curve of the form $A f_{q}^k$ to estimate state preparation and measurement error, $A$ and fidelities, $f_{q}$.  An example of experimental fidelity estimation is shown in \cref{fig:fig2}(b). Here, we exclude Paulis in $\mathcal{F}$ where fidelities are zero due to the effect of mid-circuit measurement. These zero fidelities correspond to infinite model coefficients yielding an ill-defined inverse problem, and their exclusion captures the fact that we do not attempt to learn or mitigate the information lost during the mid-circuit measurement itself. 

Measured fidelities in \cref{fig:fig2}(b) are inverted to yield values of the model coefficients in \cref{fig:fig2}(e). In general, excluding zero-fidelity Pauli terms from $\mathcal{F}$ yields a wide $M$-matrix with no unique solutions. Consistent with the application of $Z_{50\%}$ above, we additionally exclude noise terms in $\mathcal{K}$ that correspond to ancilla phase errors in the post-measurement state. Due to the absence of post-measurement ancilla phase error after full dephasing and excluding zero-fidelity Pauli terms in $\mathcal{F}$ collectively ensures that $M$ is square and full rank.  Thus the solutions are uniquely obtained through a non-negative least squares fit. For example, for a single measured qubit, we exclude $\lambda_{Y},\lambda_{Z}$ since phase errors are absent in the post-measurement state. The fidelities $f_{X},f_{Y}$ are omitted since they are trivially zero, leading to a singleton matrix $M=[1]$. An example for two qubits is shown in \cref{fig:fig2}(e).

We perform a full characterization of the mitigation procedure using mitigation validation experiments in \cref{fig:fig2}(c),(d),(f). The learning circuits are modified such that a mitigation layer (orange) is added before each twirled layer (blue) yielding \cref{fig:fig2}(c). Using this circuit, we scan the performance of the mitigated layer as a function of $k$ repeated applications of the channel. One anticipates that unity fidelities are obtained if the model coefficients are correctly learned under ideal mitigation. Experimental results are shown in \cref{fig:fig2}(d). Under repeated application of the channel, any errors  arising from a lack of sufficient samples for mitigation grow exponentially as $\gamma^{2k}$. To address this issue, the number of random circuit samples is increased exponentially with $k$ for \cref{fig:fig2}(d). Reduction of infidelities from unmitigated experiments (blue, grey) to error-mitigated experiments (orange) provides evidence for the efficacy of our procedure in \cref{fig:fig2}(f).

\begin{figure}
    \centering
    \includegraphics[scale=0.79]{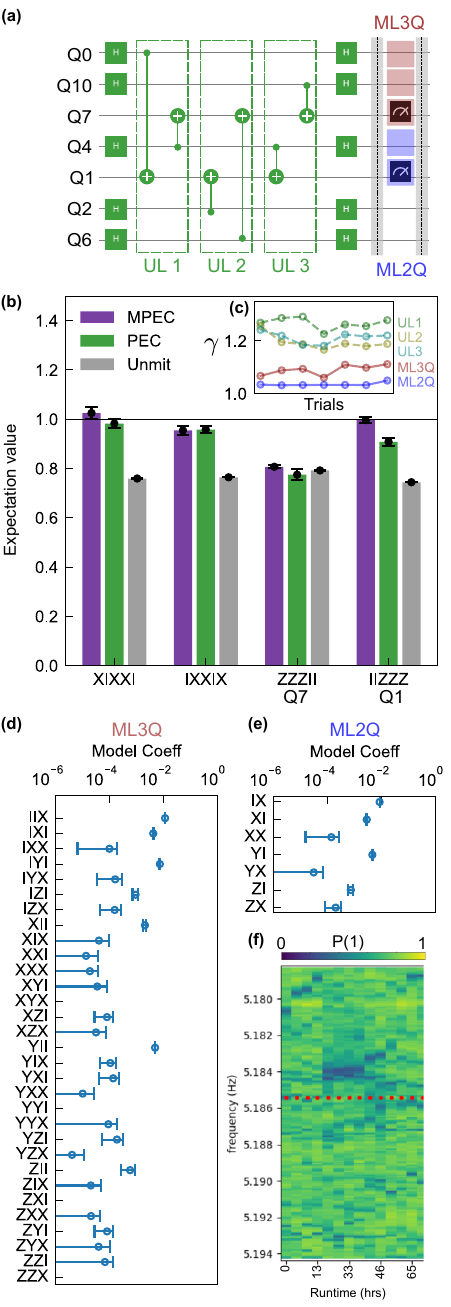}
    \caption{(a) A Clifford circuit with unitary (green) and non-unitary (purple) PEC layers on $\mathrm{ibm\_sapporo}$; Q1,7 are measured mid-circuit via alignment barriers (black lines). Circuits are dynamically decoupled (DD); if applicable, DD pulses are applied within PEC layers. For MPEC, readout topology is monitored for $12$hrs to define Q4,1 and Q0,10,7 as measurement-layers; Q2,6 are unmitigated. (b) Weight-3 stabilizers in the $X,Z$ bases are measured by mitigating all layers (purple); only unitary PEC layers (green) and an unmitigated but twirled circuit (gray) using $\sim 10$K samples collected in 7 trials over 2 days. Recovery $X$-operations for ancilla-controlled $Z$-parity checks are implemented in software. (c) $\gamma$ vs. trial index for PEC layers. (d),(e)  Learned model coefficients for Q0,10,7 and Q4,1 respectively. (f) Excited state population $P(1)$ measured by a frequency-sweep of a Stark-shifted T1 experiment versus time \cite{carroll_dynamics_2022}; dark blue regions near Q7 frequency (red dots) show a persistent TLS near Q7. Error bars show bootstrapped std. dev. for all mitigation data in (b) and a randomly chosen learning run in (d),(e).}
    \label{fig:fig3}
\end{figure}

\section{Demonstrations \label{sec:demo}}

We now extend the application of our protocols to circuits with both, unitary and non-unitary layers. Our first experiment represents a foundational tile of the $2$D surface code, as shown in \cref{fig:fig3}(a). These types of circuits have the same features as sophisticated measurement-based circuits that appear, for example, in topological quantum computing and many-body quantum simulations, where it is difficult to efficiently prepare ground states of quantum systems via unitary operations alone. Our demonstration represents a `scaled-down' version of circuits relevant to investigating ground states of a perturbed 2D surface code on a heavy-hex topology. In \cref{fig:fig3}(a), the $X$ stabilizer of the surface code is satisfied by preparing all data qubits in the $|+\rangle$ state, and two $Z$ parity checks are performed. This tile has the advantageous property that site and plaquette operators are weight-3 and symmetric, and recovery operations consists of at most two feedforward single-qubit Paulis. The ideal $\hat{\mathcal{O}}$ expectation value can be computed in software, where the feedforward operation can be implemented entirely as a sign-change in post-processing in lieu of dynamic real-time control on hardware. That is,
\begin{align}
   \langle \mathcal{O} \rangle &:= \mathrm{Tr}[\hat{\mathcal{O}}\xi(\rho)] 
    = \sum_{m} \mathrm{Tr}[{C^\dagger}^{(m)}\hat{\mathcal{O}}C^{(m)} \Pi_m \rho \Pi_m ], \nonumber \\
    &= \sum_{m} (-1)^{\langle C^{(m)}, \hat{\mathcal{O}} \rangle_{sp}} \mathrm{Tr}[\hat{\mathcal{O}} \Pi_m \rho \Pi_m ]. 
\end{align} Here $\langle C^{(m)}, \hat{\mathcal{O}} \rangle_{sp} = 1$ if the observable $\hat{\mathcal{O}}$ and the adaptive operation $C^{(m)}$ anti-commute and  zero otherwise.

Using our target circuit we compare the expectation value of the stabilizer generators under the PEC framework. The readout topology was monitored in a manner similar to \cref{fig:fig1} prior to commencing these experiments on $\mathrm{ibm\_sapporo}$. This monitoring data was used to define Q0,10,7 and Q4,1 as two measurement layers; meanwhile Q2,6 are not addressed within the PEC framework but are dynamically decoupled in all cases. These device monitoring activities are summarised in \cref{app:devicedata}. The noise learning reveals that $\gamma$-values for unitary layers dominate those of measurement-based layers in \cref{fig:fig3}(c), where higher values of $\gamma$ indicate greater intrinsic noise strength. For measurement-based layers these learned model coefficients are shown for a single learning run in \cref{fig:fig3} (d),(e). Notably, it is seen that for a 3Q measurement layer, all weight-3 noise model coefficients are nearly zero, suggesting that restricting the noise model to low-weight generators might be viable for future experiments with similar measurement amplitudes. 

Next, we test the efficacy of the learned noise model and protocols for the circuit described in \cref{fig:fig3}(a). Barring one observable, the Q7-controlled $Z$-stabilizer, the other stabilizers show good agreement with the ideal result after MPEC [\cref{fig:fig3}(b)]. In \cref{fig:fig3}(f) we observe that Q7 is subject to a persistent interaction with a two-level system (TLS) during the period of data collection. From monitoring experiments reported in \cref{app:supportingdata:fig3}, the gradual development of an interaction between an ancilla qubit and a stray TLS appears to correlate with the deteriorating performance differential between mitigated and raw expectation values of the ancilla-controlled $Z$-stabilizer. We speculate that TLS-induced readout error \cite{thorbeck2023readoutinduced} increases the probability of incorrect recovery operations applied on the basis of ancilla measurements. Noting that all two-qubit ancilla outcomes are equiprobable for the target circuit, we post-select on the `00' outcome observing some improvement in the mitigated expectation value of the $Z$-stabilizer by discarding 75\% of samples. These diagnostic observations suggest that the effects of TLS interactions are not fully captured by the learned noise model.

\begin{figure}
    \centering
    \includegraphics[scale=0.9]{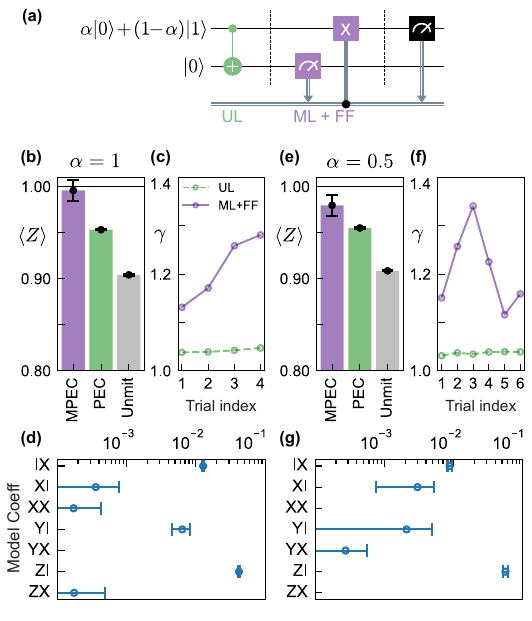}
    \caption{Error mitigated feedforward (FF) operations on $\mathrm{ibm\_peekskill}$ qubits Q10,7 where Q7 is measured; circuits are not dynamically decoupled. (a) Minimal target circuit comprising of a CNOT layer and hardware feedforward initialized with $\alpha=1$ for (b)-(d)  ($\alpha=0.5$ for (e)-(g)) on the left (right) column; experiments in columns run sequentially. (b),(e) Expectation value  $\langle Z \rangle$ when PEC mitigation is applied to all layers (purple); only unitary layers mitigated (green), or the unmitigated but twirled circuit (gray). (c),(f) $\gamma$ vs. trial index for unitary (UL) and measurement-based layer (ML) with feedforward for interleaved learning and mitigation trials. (d),(g) Learned model coefficients for a single learning instance. All FF operations incur an unaddressable communication delay of $\sim 700$ns; a dominating phase $ZI$ error on the data qubit is observed in this case. Error bars represent bootstrapped standard deviation using 2400 (3900) random circuit instances, 128 shots per circuit for $|0\rangle$ ($|+\rangle$) initial state.} 
    \label{fig:fig4}
\end{figure}

Finally, we combine measurement-based PEC with hardware feedforward \cite{gupta_encoding_2023} in \cref{fig:fig4}
employing Q10,7 on $\mathrm{ibm\_peekskill}$.
We consider two different target circuits  during the mitigation step parameterized by $\alpha$ in \cref{fig:fig4}(a). With $\alpha=1$ 
in \cref{fig:fig4}(b)-(d), the CNOT application is trivial, and no classically-controlled $X$ operation is applied in the ideal case. If instead the initial state is prepared  with $\alpha=0.5$ in \cref{fig:fig4}(e)-(g), the state on the data qubit is susceptible to phase error and  ancilla qubit outcomes are equiprobable resulting in non-trivial recovery operations. For all mitigation experiments, the learning procedure is identical and unaffected by the choice of $\alpha$.

In \cref{fig:fig4}, the measurement-based PEC protocol (purple) outperforms all other experiments (green, gray) as depicted by panels (b),(e). This performance differential is further contextualized with a higher $\gamma$ for the non-unitary layer in \cref{fig:fig4}(c),(f). The learned model coefficients in a single trial  \cref{fig:fig4}(d),(g) are dominated by phase error on the data qubits, $ZI$, confirming our sanity-check that phase shifts and dephasing noise during real-time control delays are twirled to yield phase errors that dominate the channel. Noting that the data-qubit is projected into an eigenstate of the $Z$-operator after the ancilla is measured, the state of the data-qubit after mid-circuit measurement is immune to phase errors introduced during communication with the real-time controller. This observation suggests that the performance gains accrued under MPEC cannot be attained simply through mitigating idling error during real-time communication.

The inclusion of mid-circuit measurements and feedforward raises the possibility that classical errors may arise on the classical wire after measurement. As an extreme example, readout-clouds for different states of the qubit may be well-separated \cite{magesan2015machine,gambetta2007protocols} but rotated relative to a linear discriminator applied in software, leading to a classical state discrimination error. Such errors are purely classical in origin and therefore they are not amplified and learned in our procedure. Nevertheless these errors may affect the correctness with which recovery operations are applied in hardware and could account for the offset from the ideal value of unity in \cref{fig:fig4}(e) under MPEC. Signatures of classical errors on the control wire are discussed further in \cref{app:supportingdata:fig4} and addressing this type of classical discriminator error is the subject of future work.

\section{Conclusion \label{sec:conclusion}}

Our work establishes the PEC framework as a unified approach for error-mitigation for measurement-based circuits. We demonstrate a scalable, experimentally viable procedure for learning noise in measurement-based circuits. We find that by appropriately characterizing the readout topology relevant for measurement-based PEC on a particular device, the learning procedure remains scalable while also taking into account non-nearest-neighbor interactions. Our target demonstrations test the efficacy of our protocols, and in conjunction with device monitoring experiments, provide an insight into observed non-idealities. Our techniques present a valuable contribution to realising benefits of dynamic, measurement-based quantum computation.

\section{Code and data availability}
The codebase and data used in this study is available on reasonable request. 


\section{Acknowledgements}
Research on measurement model learning was partially sponsored by the Army Research Office and was accomplished under Grant Number W911NF-21-1-0002. The views and conclusions contained in this document are those of the authors and should not be interpreted as representing the official policies, either expressed or implied, of the Army Research Office or the U.S. Government. The U.S. Government is authorized to reproduce and distribute reprints for Government purposes notwithstanding any copyright notation herein. R.S.G is grateful to Neereja Sundaresan, Holger Haas, Youngseok Kim  and Thomas Alexander for experimental guidance and device performance management; Ted Thorbeck and Luke Govia for useful discussions on understanding measurements, Ben Brown for developing a small surface code demonstration, Brad Mitchell and Andrew Eddins for ongoing error-mitigation discussions, and Ted Thorbeck, Andrew Eddins and Luke Govia for feedback on the manuscript. We also acknowledge the work of the IBM Quantum software and hardware teams that enabled this project.

\clearpage
\appendix
\onecolumngrid

\section{ Supporting theoretical analysis for measurement-based PEC  \label{app:tmbc}}

\subsection{Definitions \label{app:tmbc:definitions}}

We establish notation and definitions in this section for the main text and the appendix. As noted in the main text, the ideal map $\xi$ defined on a bipartite system describing unmeasured data qubits  and measured ancilla qubits is defined as,
\begin{align}
    \xi (\rho) := \Sigma_{m \in \mathcal{M}} C^{(m)} \Pi_{(m)} \rho  \Pi_{(m)}^\dagger
 C^{(m)}{}^\dagger, 
\end{align} where $m$ is the  outcome of the projectively measuring the ancilla qubit, $\mathcal{M}$ is the set of all possible outcomes on ancilla qubits i.e., $\mathcal{M}:= \{0, 1\}$ for a single qubit. The projector $ \Pi_{(m)}$ labels the state of the system after an ancilla qubit measurement, $\Pi_{(m)} \rho  \Pi_{(m)}^\dagger$, and $C^{(m)}$ is a Pauli feedforward operation triggered by the outcome $m$. For clarity, both operations are defined on data and ancilla qubits,
\begin{align}
    \Pi_{(m)} &:= I \otimes |m \rangle \langle m |,  \\
    C^{(m)} &:= C_D^{(m)} \otimes C_A^{(m)}. 
\end{align}  

As discussed in the main text, we enforce that all phase information is lost on a post-measurement state by adding a $Z$ with probability $50$\% immediately after each mid-circuit measurement. This action enforces that a strongly projective measurement occurs in the computational basis and completely dephases the post measurement state yielding $\Pi_{m} \rho \Pi_{m}^\dagger$. Defining some intrinsic noise, $\tilde{\Lambda}$, and applying a $Z_{50\%}$ directly after measurement, one writes the noisy measurement-based channel $\tilde{\xi}$ as,
\begin{align}
    \tilde{\xi} := \xi \circ \tilde{\Lambda}.
\end{align}
Performing Pauli twirling on a channel $\chi$ is defined by the operation,
\begin{align}
 \mathrm{Tw}[\chi]:= \mathbb{E}[ P_i^\dagger\chi(P_i \cdot P_i^\dagger)P_i ]_i,
\end{align} where $P_i$ represent elements of the Pauli group indexed by $i$, and $\mathbb{E}$ denotes an averaging over these elements. For a bipartite system of data and ancilla qubits, we use super-scripts to denote individual Paulis in each sub-system, for example,
\begin{align}
    P_i &:= P_i^{(D)} \otimes P_i^{(A)}.
\end{align} The Pauli transfer matrix of any channel $\chi$ shows how Pauli coefficients of a $n$-qubit state transforms under the channel, with matrix elements given by
\begin{align}
    R^{\chi}_{a,b} := \frac{1}{4^n} \mathrm{Tr}\left[ P_a \chi(P_b). 
\right]\end{align} An additional function that will be useful is the classical bitwise negation of the projective measurement outcome $m$ according to some condition $x$,
\begin{align}
\eta (m | x) := \begin{cases} m \text{ if } x \\ \neg m , \text{ if } \mathrm{not}(x) \end{cases}.
\end{align} This negation function will be used to describe both ideal operations and stochastic functions on the classical wire associated with the raw measurement outcome $m$. Finally the notation $\delta_{(a,b)}$ is the Kronecker delta function which is unity if $a=b$ and zero otherwise.

\subsection{Twirling measurements \label{app:tmbc:twirlingmsmts}}

We now describe how to push a twirl through a projective measurement in the computational basis. 

Suppose we sample $P_i$ uniformly at random and apply it to all qubits before a mid-circuit measurement $\Pi_{m}$ is performed. If $P_i^{(A)} $ on a single ancilla qubit  before measurement is either $I$ or a $Z$, then the same twirl can be applied after the measurement as all phase information is lost. Alternatively, if $P_i^{(A)}$ is either $X$ or a $Y$  before measurement, then again applying $P_i$  after measurement undoes the twirl on the measured quantum wires, assuming all phase information is lost perfectly. However, the ancilla outcome $m$ also needs to be flipped and the twirl-adjusted classical outcome is denoted by $\neg m$. Both of these cases can be summarized by using $\eta (m | x)$ to define a function $f(m, P)$,
\begin{align}
    f(m, P) := \eta(m | P \in \{I, Z\}),
\end{align}  and it is useful to note that $\neg f(m, P) = f(\neg m, P) $ for single qubits. Furthermore, the  properties
\begin{align}
    X |m\rangle \langle m | X  &= |\neg m\rangle \langle \neg m |, \\
    Z |m\rangle \langle m | Z &=  |m\rangle \langle m |, 
\end{align} imply that a term of the form $P_i \Pi_{f(m,  P_i^{(A)})} P_i $ reduces to the  projector $\Pi_m$, 
\begin{align}
    P_i \Pi_{f(m,  P_i^{(A)})} P_i & = P_i \Pi_{\eta(m|  P_i^{(A)} \in \{ I, Z\})} P_i \\
    &= \begin{cases}  I \otimes P_i^{(A)} |m\rangle \langle m | P_i^{(A)},  &\quad P_i^{(A)} \in \{I, Z\}  \\
                                         I \otimes P_i^{(A)} |\neg m\rangle \langle \neg  m | P_i^{(A)} ,  &\quad P_i^{(A)} \in \{X, Y\} \end{cases} \\
    &= \begin{cases} \Pi_{(m)}, & \quad P_i^{(A)} \in \{I, Z\}  \\
                                \Pi_{\neg(\neg m)},  &\quad P_i^{(A)} \in \{X, Y\} \end{cases} \\
    &= \Pi_{(m)} , \forall P_i.
\end{align} Hence the twirl $P_i$ is pushed through the measurement projector,
\begin{align}
     P_i \Pi_{f(m,  P_i^{(A)})}= \Pi_{(m)} P_i.
\end{align} With this notation, one can ask whether a Pauli feedforward operation $C$ is triggered by twirl-adjusted measurement outcome $f(m, P_i^{(A)})$ or the original measurement outcome. One pushes a Pauli twirl through the triggering condition in notation as,
\begin{align}
  & P_i C^{f(m,  P_i^{(A)})} \Pi_{f(m,  P_i^{(A)})} =  \left( P_i C^{f(m,  P_i^{(A)})}  P_i \right)  P_i \Pi_{f(m,  P_i^{(A)})} = \left( P_i C^{(m)}  P_i \right) \Pi_{(m)} P_i, 
\end{align} where the twirl is pushed through the measurement and the triggering condition on the feedforward operation is adjusted accordingly in the last step.  Conjugating a Pauli feedforward operation with Pauli twirls gives
\begin{align}
    P_i C^{(m)} P_i = (-1)^{\langle C^{(m)}, P_i \rangle_{sp}} C^{(m)} \label{eqn:twirlfeedforward}
\end{align} where $\langle A,B \rangle_{sp}$ denotes the symplectic inner product which is $0$ if $A,B$ commute and $1$ if they anticommute. Thus pushing a Pauli twirl through measurement and Pauli feedforward is given by,
\begin{align}
    P_i C^{f(m,  P_i^{(A)})} \Pi_{f(m,  P_i^{(A)})} = (-1)^{\langle C^{(m)}, P_i \rangle_{sp}} C^{(m)} \Pi_{(m)} P_i. \label{eqn:pushtwirlthroughffwd}
\end{align}

Classical errors on the control-wire may transform the true measurement outcomes during classical processing (e.g., an outcome subject to state-independent and symmetric discriminator error with error probability $r \in [0, 0.5)$). We unify these choices by writing $C^{g(m, P_i)}$, where $g(m, P_i)$ is a general classical outcome to trigger classical control for each twirl element $P_i$, 
\begin{align}
    g(m, P_i^{(A)}) :&= \eta(f(m, P_i^{(A)})| u < 1-r), \quad u \sim \mathrm{Uniform}(0,1) \\
    &= \begin{cases} 
    f(m, P_i^{(A)}), &\text{ with prob. $1-r$} \\ 
    \neg f(m, P_i^{(A)}), & \text{ with prob. $r$.}
    \end{cases} \\
    &= \begin{cases} 
    f(m, P_i^{(A)}), &\text{ with prob. $1-r$} \\ 
     f(\neg m, P_i^{(A)}), & \text{ with prob. $r$.}
    \end{cases} \\
    &= f(\eta(m| u < 1-r), P_i^{(A)}), \quad u \sim \mathrm{Uniform}(0,1), 
\end{align} where $u$ is sampled uniformly for each instance. One notes that  $C^{f(\eta(m| u < 1-r), P_i^{(A)})} \to C^{f(m, P_i^{(A)})}$ for no discriminator error $r \to 0$. 

With this in mind, we can perform a twirling operation on the noisy measurement channel as follows,
\begin{align}
    \mathrm{Tw}[\tilde{\xi}] &= \mathbb{E}[ P_i^\dagger\tilde{\xi}(P_i \cdot P_i^\dagger)P_i ]_i = \mathbb{E}[ P_i^\dagger( \xi \circ \tilde{\Lambda})(P_i \cdot P_i^\dagger)P_i ]_i, \\
    &=\Sigma_{m \in \mathcal{M}} \mathbb{E}\left[ P_i^\dagger C^{g(m,  P_i^{(A)})} \Pi_{f(m, P_i^{(A)})}  \tilde{\Lambda}(P_i \cdot P_i^\dagger)  \Pi_{f(m, P_i^{(A)})}^\dagger C^{g(m,  P_i^{(A)})}{}^\dagger P_i \right]_i.
\end{align} Using \cref{eqn:pushtwirlthroughffwd}, we push the twirl through the measurement as well as the triggering condition for feedforward, 
\begin{align}
   \mathrm{Tw}[\tilde{\xi}] &=\Sigma_{m \in \mathcal{M}} \mathbb{E}\left[ P_i^\dagger C^{f(\eta(m| u < 1-r), P_i^{(A)})} \Pi_{f(m, P_i^{(A)})}  \tilde{\Lambda}(P_i \cdot P_i^\dagger)  \Pi_{f(m, P_i^{(A)})}^\dagger C^{f(\eta(m| u < 1-r), P_i^{(A)})}{}^\dagger P_i \right]_i,\\
   &=\Sigma_{m \in \mathcal{M}} \mathbb{E}\left[ C^{\eta(m| u < 1-r)} \Pi_{(m)} P_i^\dagger \tilde{\Lambda}(P_i \cdot P_i^\dagger)  P_i \Pi_{(m)}^\dagger C^{\eta(m| u < 1-r)}{}^\dagger \right]_i, 
\end{align} where Pauli twirls are pushed through the triggering condition as well as the measurement; the signs cancel and the superscripts on feedforward operations capture classical errors on the control wire. The final form of the twirled channel is,
\begin{align}
    \mathrm{Tw}[\tilde{\xi}] &=\Sigma_{m \in \mathcal{M}} \mathbb{E}\left[ C^{\eta(m| u < 1-r)} \Pi_{(m)} P_i^\dagger \tilde{\Lambda}(P_i \cdot P_i^\dagger) P_i \Pi_{(m)}^\dagger  C^{\eta(m| u < 1-r)} \right]_i, \\
    &=\Sigma_{m \in \mathcal{M}}  C^{\eta(m| u < 1-r)} \Pi_{(m)} \mathbb{E}\left[ P_i^\dagger \tilde{\Lambda}(P_i \cdot P_i^\dagger)  P_i \right]_i \Pi_{(m)}^\dagger  C^{\eta(m| u < 1-r)}, \label{eqn:twirledchannel}\\
    &\equiv \xi \circ \Lambda,
\end{align} where $\Lambda := \mathbb{E} \left[  P_i^\dagger \tilde{\Lambda}(P_i \cdot P_i^\dagger) P_i  \right]_i$ is the twirled noise.
\\
\\
While the noise is diagonalized, the full Pauli transfer matrix of the ideal channel and noise $ \xi \circ \Lambda$ is not diagonal, irrespective of discriminator error rate $r$. We assume that there is no feedforward operation on the ancilla qubits, yielding,
\begin{align}
    R^{\mathrm{Tw}[\tilde{\xi}]} &= R^{\xi}  R^{\Lambda}\\
    R^{\Lambda}_{a,b} :&= f_a \delta_{a, b} \\
    R^{\xi}_{a,b} :&= \frac{1}{d} \Sigma_m  \mathrm{Tr}\left[  P_a  C^{\eta(m| u < 1-r)}\Pi_{(m)}P_b \Pi_{(m)}^\dagger C^{\eta(m| u < 1-r)}\right],\\
    &= \frac{1}{d} \Sigma_m  \mathrm{Tr}\left[ (-1)^{\langle C^{\eta(m| u < 1-r)}, P_a^{(D)}\rangle_{sp} }  P_a\Pi_{(m)}P_b \Pi_{(m)}^\dagger \right], \\
    &=  \delta_{(P_a^{(D)},P_b^{(D)})} \Sigma_m (-1)^{ \langle C^{\eta(m| u < 1-r)}, P_a^{(D)}\rangle_{sp} }  \langle m| P_a^{(A)}  | m\rangle \langle m| P_b^{(A)}|m\rangle ,
\end{align} where $\mathrm{Tr}\left[  I_{n_A} \right] \mathrm{Tr}\left[  P_a^{(D)}P_b^{(D)} \right]/d = \delta_{(P_a^{(D)},P_b^{(D)})} $. Meanwhile the terms $\langle m| P^{(A)}  | m\rangle$ are nonzero only for,
\begin{align}
    \langle m| I  | m\rangle &= 1,  \forall m \\
    \langle m| Z  | m\rangle &= (-1)^m. 
\end{align} Using the relations above, and the (data, ancilla) basis ordering $II, IZ, XI, XZ, YI, YZ, ZI, ZZ, IX, IY,  \hdots$, we find that the top quadrant of the Pauli transfer matrix (PTM) has non-zero values and is block diagonal with the following $2\times2$ block for each $m$,
\begin{align}
    (-1)^{\langle C^{\eta(m| u < 1-r)}, P_j^{(D)}\rangle_{sp}} \begin{bmatrix}
    1 & (-1)^m \\ (-1)^m & 1
    \end{bmatrix}, 
\end{align} appearing along the diagonal, where $j$ indexes Paulis on the data qubit, and zero elsewhere. The summation of these block-diagonal regions over $m$ will eliminate off-diagonal elements when the feedforward operations commute with the data-qubit Pauli basis $j$, and non-zero diagonal elements will be retained for the anticommuting case. \\
\\

\subsection{Identity feedforward \label{app:tmbc:idffwd}}
We will always have a diagonalized Pauli transfer matrix if $C^{\eta(m| u < 1-r)} \equiv I, \quad  \forall  m \in \mathcal{M} $. In this case of identity feedforward, 
 \begin{align}
    \mathrm{Tw}[\tilde{\xi_{I}}] 
    &=\Sigma_{m \in \mathcal{M}} \Pi_{(m)} \mathbb{E} \left[  P_i^\dagger \tilde{\Lambda}(P_i \cdot P_i^\dagger) P_i  \right]_i \Pi_{(m)}^\dagger 
    = \xi_{I} \circ \Lambda \label{eqn:twirledchannelidffwd}.
\end{align} The transfer matrix $R^{\tilde{\xi_{I}}} = R^{\xi_{I}}  R^{\Lambda}$ simplifies to,
\begin{align}
    R^{\xi_{I}}_{a,b} =  \delta_{(P_a^{(D)},P_b^{(D)})} \Sigma_m \langle m| P_a^{(A)}  | m\rangle \langle m| P_b^{(A)}|m\rangle ,
\end{align} where the sum over $m$ eliminates the off-diagonal terms,
\begin{align}
    \Sigma_m \langle m| P_a^{(A)}  | m\rangle \langle m| P_b^{(A)}|m\rangle = \delta_{(P_a^{(A)},P_b^{(A)})}, \quad \forall P_a^{(A)},P_b^{(A)} \in \{I, Z\},  
\end{align} giving a diagonal PTM with seven non-zero Pauli fidelities in the two-qubit case, excluding identity, 
\begin{align}
     R^{\tilde{\xi}_{I}}_{a,b} =  f_a \delta_{(a, b)} \quad \forall P_a^{(A)},P_b^{(A)} \in \{I, Z\}, \text{ else $0$.}
\end{align}

\subsection{Pauli feedforward \label{app:tmbc:pauliffwd}}

In using feed-forward operations, we note that the dominant source of noise accrues as coherent phase errors and/or dephasing on idling qubits due to classical communication delays in the control flow. Once the control decision from the real-time controller has been returned, any conditional single-qubit gates are effectively noiseless to implement in comparison with other error rates in the circuit.\\
\\
All learning benchmarking circuits are thus run by using a conditional delay operation. By scheduling a conditional delay operation with a duration matching that of the single-qubit gate, we can trigger the actual real-time control path accruing additional  idling times of $500-900$ns in the circuit. This noise channel has a diagonalized Pauli transfer matrix identical to that in section \cref{app:tmbc:idffwd}.\\
\\
During mitigation, the original non-identity Pauli feedforward operations are used in the mitigation layer but with no changes to the twirling procedures.  Since mitigation layers are applied before the noise channel, we see that the Paulis sampled from the inverse distribution, denoted by $\rho_M$, act as an input to the diagonalized intrinsic noise in \cref{eqn:twirledchannel},
\begin{align}
    \Sigma_{m \in \mathcal{M}}  C^{\eta(m| u < 1-r)} \Pi_{(m)} \mathbb{E}\left[ P_i^\dagger \tilde{\Lambda}(P_i \rho_M P_i^\dagger)  P_i \right]_i \Pi_{(m)}^\dagger  C^{\eta(m| u < 1-r)}.
\end{align}  This mitigation procedure for non-identity feedforward is verified in experiment in Fig.~4 of the main text.

\begin{figure}
    \centering
    \includegraphics[scale=0.8]{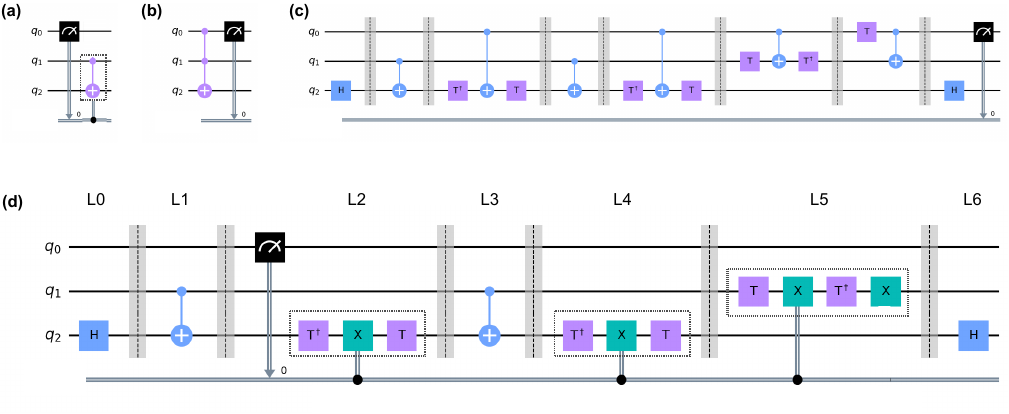}
    \caption{(a) A classically-controlled CNOT gate. (b) Representation of a Toffoli gate followed by a measurement on a control wire, using the principle of deferred measurements. (c) Unitary decomposition of Toffoli gate using 6 CNOTs \cite{shende_cnot-cost_2009}. (d) Resulting measurement-based decomposition of a classically-controlled CNOT by pushing measurement through the Toffoli gate, yielding a control qubit wire and secondary control via single classical register. Dashed boxes denote that the classically-controlled single-qubit gates $T^\dagger XT$ and $TXT^\dagger X$ are Clifford and the layers $L4$ and $L5$ can be parallelized.
    \label{app:ToffoliDecomp}}
\end{figure}

\subsection{Clifford feedforward \label{app:tmbc:cliffordffwd}}

In this section we analyse whether any classically-controlled Clifford circuit is permitted by our procedure. Any Clifford feedforward circuit can be decomposed into classically-controlled single-qubit Clifford and CNOT gates. However, one concern is that the inclusion of classically controlled CNOT gates in Clifford feedforward circuits could lead to an exponential growth in the number of nested, classically-controlled two-qubit gates that are unfeasible to implement. 

We propose that it is possible to replace a classically controlled CNOT with a measurement-based circuit of increased depth as shown in \cref{app:ToffoliDecomp}. In particular, a classically controlled CNOT is equivalent to a Toffoli gate with a measurement on one of the control wires, as shown in panels (a),(b). It is well known that the three-qubit Toffoli gate can be decomposed in terms of five two-qubit gates, and this number increases to six if two-qubit gates are restricted to CNOTS \cite{barenco_elementary_1995,ralph_efficient_2007,shende_cnot-cost_2009}.

For the given decomposition of a Toffoli gate \cite{shende_cnot-cost_2009} in \cref{app:ToffoliDecomp}(c), one can push the measurement on the first control wire through the unitary circuit yielding \cref{app:ToffoliDecomp}(d). Noting that a $T$-gate followed by measurement will not affect the probability of measurement outcomes and can be safely dropped, any CNOT controlled by the first wire can be converted to a classically controlled $X$ gate. We confirm that the single-qubit $T^\dagger X T$ and $T X T^\dagger X$ are Clifford.

Thus to accommodate classically-controlled Clifford circuits, MPEC only needs to accommodate single-qubit Clifford operations. Instead of applying the same Pauli $P_i$ on both sides of the feedforward operation $C$ in \cref{eqn:twirlfeedforward}, a conjugate Pauli $Q_i$ is computed straightforwardly up to phase, $Q_i C = C P_i$. Thus single-qubit classically-controlled Clifford feedforward operations are learned with identity-feedforward, followed by mitigation with Clifford feedforward with the appropriate computation of the conjugate Pauli twirl. Experimentally verifying the error mitigation efficacy of MPEC for classically-controlled Clifford circuits is the subject of future work.

\section{ Device characterization  \label{app:devicedata}} 

All experimental data in the main text was collected on IBM Falcon processors \cite{ibm_processortypes_2023}. Characterization data for all devices used for figures of the main text are summarized in Tables~\ref{table:device},\ref{table:cr}. Data collection for Fig.~1 - Fig.~3 spanned a total of two weeks on  $\mathrm{ibm\_sapporo}$ r5.11 over a single region outlined in Fig.~1 of the main text. During this time, monitoring experiments were interleaved with data collection trials. Device coherence times for  $\mathrm{ibm\_sapporo}$ lie in the range $\sim 50-80 \mu$s. Meanwhile CX gates, constructed from echoed cross-resonance pulse sequence, are specified in one direction, with the reverse directions accessed by addition of single qubit gates. Error per gate (EPG) is extracted from isolated two-qubit randomized benchmarking (spectator qubits idling) and is found to range from 0.97-1.44\%. For Fig.~4, data was collected on $\mathrm{ibm\_peekskill}$ r8  with dynamic circuits. 

\begin{table*}[h]
\tabcolsep=0.2cm
\begin{tabular}{|l|c|c|c|c|c|c|c|c|c|}
\hline
Device & Qubit &  Qubit freq. &  Anharmonicity  &  T1  &  T2 echo &       EPG &  RO fid. &  P(0$\vert$1) &  P(1$\vert$0) \\
  &   &   (GHz) & (MHz) &  (us) &  (us) &  &  (\%) &  &   \\
\hline
$\mathrm{ibm\_sapporo}$ & 0     &             5.0649 &               -342.5 &     60.2 &          24.7 &  0.000408 &         99.1 &   0.012 &   0.006 \\
$\mathrm{ibm\_sapporo}$ & 1     &             5.2202 &               -340.8 &     72.1 &          40.3 &  0.000385 &         98.8 &   0.015 &   0.009 \\
$\mathrm{ibm\_sapporo}$ & 10    &             5.1142 &               -341.8 &     59.4 &          95.2 &  0.000469 &         98.9 &   0.014 &   0.007 \\
$\mathrm{ibm\_sapporo}$ & 2     &             5.0945 &               -342.3 &     71.3 &         100.2 &  0.000338 &         98.6 &   0.017 &   0.011 \\
$\mathrm{ibm\_sapporo}$ & 4     &             5.3505 &               -337.8 &     60.6 &         100.7 &  0.000304 &         98.6 &   0.016 &   0.012 \\
$\mathrm{ibm\_sapporo}$ & 6     &             5.2850 &               -339.1 &     61.9 &         109.7 &  0.000345 &         98.9 &   0.013 &   0.009 \\
$\mathrm{ibm\_sapporo}$ & 7     &             5.1854 &               -340.7 &     71.4 &          94.8 &  0.000533 &         96.2 &   0.043 &   0.033 \\
\hline
$\mathrm{ibm\_peekskill}$ & 10    &           4.8376 &               -345.0 &    294.2 &         264.5 &  0.000155 &         99.0 &    0.010 &   0.009 \\
$\mathrm{ibm\_peekskill}$ & 7     &           4.7280 &               -346.7 &    315.0 &         296.7 &  0.000381 &         98.3 &    0.020 &   0.013 \\
\hline
\end{tabular}
\caption{Average single-qubit gate benchmarks on section of $\mathrm{ibm\_sapporo}$ (top panel) used in Fig.~1 - Fig.~3, and two qubits on $\mathrm{ibm\_peekskill}$ (last row) used in Fig.~4 in the main text. 
\label{table:device}} 
\end{table*}

\begin{table}

\centering
\begin{tabular}{|l|c|c|c|}
\hline
Device & Gate &  CX length (ns) &   EPG (\%) \\
\hline
$\mathrm{ibm\_sapporo}$ & 0\_1  &           412.4 &  1.440878 \\
$\mathrm{ibm\_sapporo}$ & 1\_4  &           440.9 &  1.184191 \\
$\mathrm{ibm\_sapporo}$ & 2\_1  &           376.9 &  1.311541 \\
$\mathrm{ibm\_sapporo}$ & 6\_7  &           277.3 &  0.970609 \\
$\mathrm{ibm\_sapporo}$ & 7\_10 &           341.3 &  1.044530 \\
$\mathrm{ibm\_sapporo}$ & 7\_4  &           647.1 &  1.390997 \\
\hline
$\mathrm{ibm\_peekskill}$ & 7\_10 &           455.1 &  0.520371 \\
\hline
\end{tabular}
\caption{Average two-qubit gate benchmarks on section of $\mathrm{ibm\_sapporo}$ (top panel) used in Fig.~1 - Fig.~3, and gate CX$7\_10$ of $\mathrm{ibm\_peekskill}$ used in Fig.~4 of the main text. \label{table:cr}}
\end{table}

As reported in Fig.~1, we use measurement-induced crosstalk experiments to illustrate readout topology selection for MPEC. In each measurement-induced crosstalk experiment, a pulse-sequence $X_{\pi/2}$--$ \frac{\tau}{2}$--$X_{\pi}$--$ \frac{\tau}{2}$--$Y_{\pi/2}$ is performed on the data qubit with a fixed wait time $\tau = 2\mu$s. Meanwhile a measurement pulse with a fixed measurement amplitude is applied to the ancilla qubit at the start of the delay in the sequence. The ideal probability of the data qubit in the excited state, $P(1)=0.5$, is compared to the experimentally measured probability for a sufficiently high measurement amplitude on the ancilla qubit  i.e. $2\times$ to $4\times$ the calibrated measurement amplitude.  The absolute value of this $P(1)$ error is taken is reported in Fig.~1. For capacitively coupled qubit-pairs, readout cross-talk is observed only for the pair Q10,Q7 when a high-amplitude measurement pulse is applied to Q7. Finally, it is also not essential to use measurement-induced crosstalk experiments to select an appropriate readout topology and other metrics could be considered instead e.g. see Ref.~\cite{govia_randomized_2022}.

\begin{figure}[h]
    \centering
    \includegraphics[scale=0.75]{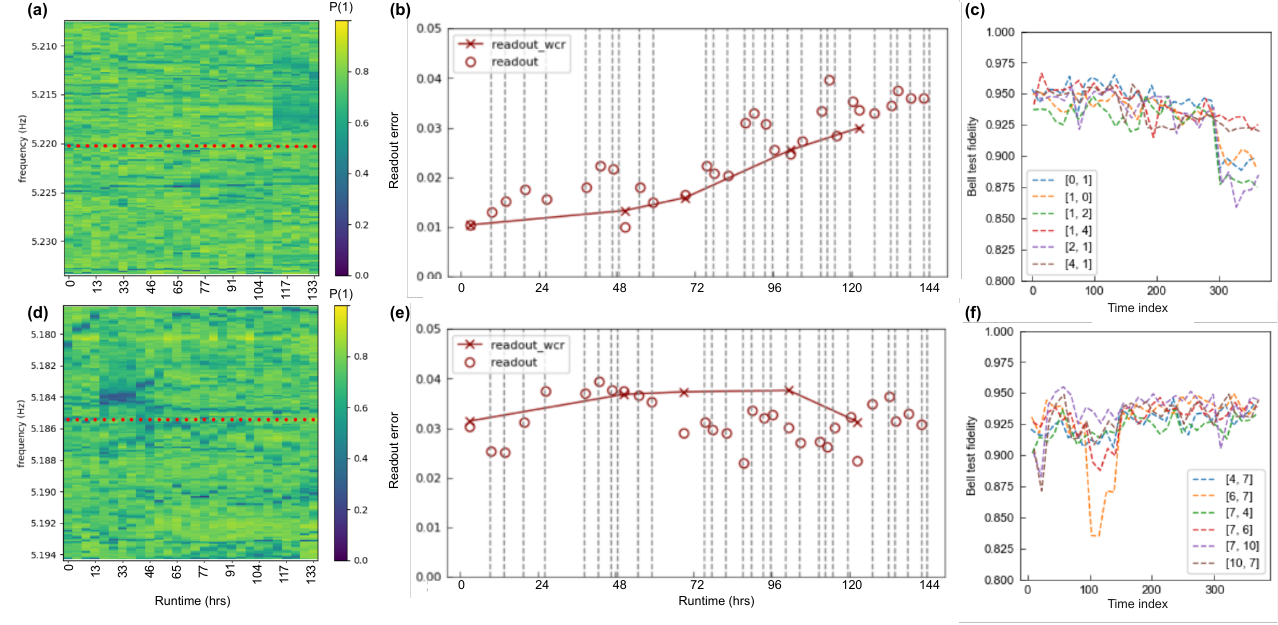}
    \caption{\label{fig:app:postselectioninfig3:monitoring}Ancilla-qubit monitoring for region of $\mathrm{ibm\_sapporo}$ where the $x$--axis of all panels represents the same monitoring period.  Data collection in Fig. 3 of the main text approximately corresponds to the first third of monitored device time in all panels of this figure. Top (bottom) rows represent data for ancilla qubit Q1 (Q7). (a),(d) Stray coupling to undesired two-level-systems (TLS) characterization by measuring $P(1)$ in a T1 experiment (color scale) vs. sweep of stark-shifted qubit frequencies ($y$-axis) repeated in approximately 12 hour intervals ($x$-axis)\cite{carroll_dynamics_2022}. Evidence of a persistent long-lived TLS is observed for Q1 in the second half of the monitoring period, and on Q7 for the entirety of monitoring period. (b),(e) Readout error where qubit initialization involves either unconditional (crosses) and conditional (circles) reset. Grey vertical dashed lines approximately denote the start time of repeated cycles of learning and mitigation where cycle can contribute a maximum of 1500 mitigation samples to the overall database. (c),(f) Hellinger fidelity for the Bell-test sequence $H_{(1)}$($\mathrm{CX}_{(1),(2)}$)${}^k$ for $k=5$ repetitions and three rounds of conditional resets during initialization, where $(i)$ is an index $i=1,2$ labeling any pair of qubits. The Bell test is repeated for pairs of qubits associated with connected edges for each ancilla qubit; edges in the same PEC layer are tested in parallel to match conditions in actual experiment.} 
\end{figure}

\begin{figure}[h]
    \centering
    \includegraphics[scale=0.8]{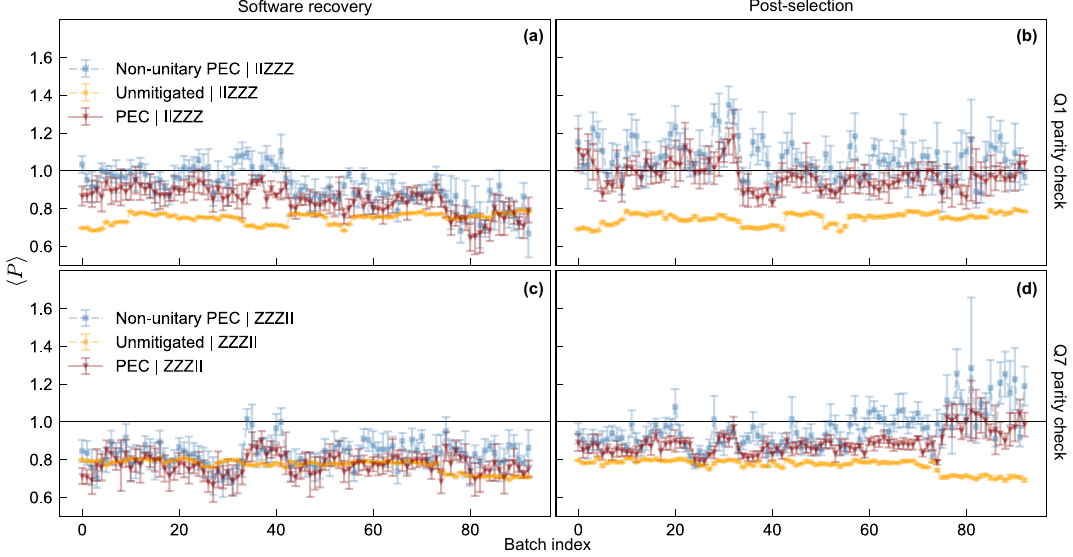}
    \caption{Expectation values from batches of 150 random circuit samples vs. temporal execution index; coloured data markers represent circuits using MPEC (blue), PEC (red) and unmitigated but twirled circuits (yellow). These experiments were interleaved with monitoring experiments in \cref{fig:app:postselectioninfig3:monitoring}. Top (bottom) rows depict $Z$-type parity checks controlled by Q1 (Q7). (a),(c) Expectation values where recovery operations are implemented in software and all circuit samples are retained. (b),(d) Expectation values after post-selection, where any random circuit sample where mid-circuit measurements on any ancilla yields a `1' outcome is discarded. Expectation values and error bars representing a single standard deviation are estimated from bootstrapping within each batch. \label{fig:app:postselectioninfig3:timeseries}}
\end{figure}

\section{Supporting analysis for Fig. 3  \label{app:supportingdata:fig3}}

\begin{figure}[h]
    \centering
    \includegraphics[scale=0.8]{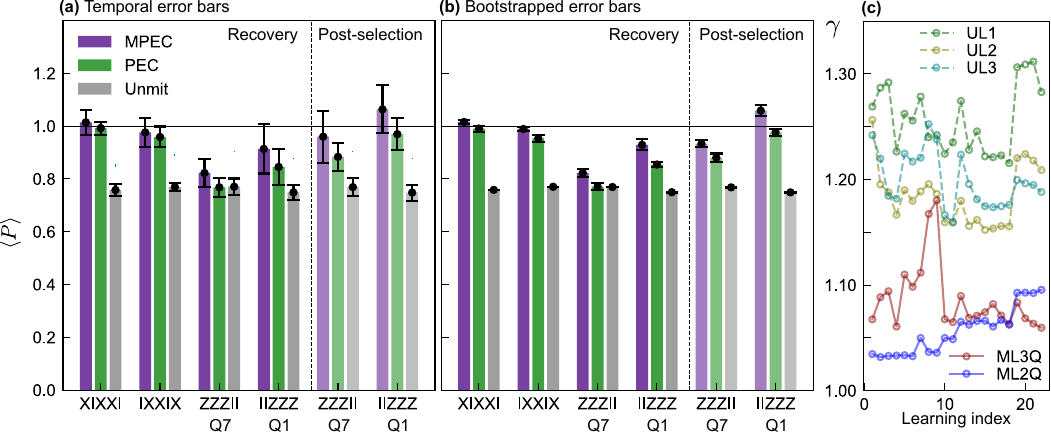}
    \caption{Stabilizer expectation value over all trials of the monitored experiment consisting of $\sim 30 $K samples collected over $\sim 22 $ trials over a period of 5 days. (a) Expectation values computed in batches of 1500 samples, where error bars represent temporal variation over subsequent trials. (b) Expectation values bootstrapped using the full database where error bars represent an estimate of standard deviation from bootstrapping. In both (a) and (b), a vertical dashed line separates the case where all samples are retained (left) vs. the case where samples are post-selected on the `00' outcome on both ancilla qubits (right); subscripts on $x$-axis denote ancilla associated with each parity check. (c) $\gamma$-value per layer vs. temporal execution order for each layer.    }
    \label{fig:app:postselectioninfig3:expvals}
\end{figure}

We review the experimental considerations in collecting and analyzing data for Fig. 3 in the main text. In \cref{fig:app:postselectioninfig3:monitoring,fig:app:postselectioninfig3:timeseries,fig:app:postselectioninfig3:expvals} approximately $\sim 30$K mitigation samples were collected over $\sim  22$
trials and data collection was monitored over $\sim 5$ days. In \cref{fig:app:postselectioninfig3:monitoring} we monitor coupling of ancilla qubits to TLS defects over data collection spanning many cycles of learning and mitigation. Monitoring experiments suggest that a long-lived TLS interaction develops for Q1 in \cref{fig:app:postselectioninfig3:monitoring}(a)-(c). This interaction is signaled by the increase in readout error \cite{thorbeck2023readoutinduced} and gradual drop in Hellinger fidelity for preparing a Bell state for edges relevant to our circuit. For this reason, only the first 7 trials are reported in Fig. 3 noting that the resulting number of mitigation samples sufficiently exceeds the sampling requirement estimated using the total $\gamma$ for the circuit. Meanwhile Q7 appears to be subject to a TLS interaction for the entire period in \cref{fig:app:postselectioninfig3:monitoring}(d)-(f) and it has the highest readout error in \cref{table:device}.

In all panels of \cref{fig:app:postselectioninfig3:timeseries}, the $x$-axis represents batches of 150 mitigation samples in temporal execution order that are bootstrapped within each batch. Meanwhile the $y$-axis depicts expectation values of the ancilla-controlled $Z$ stabilizer computed from samples in each batch for Q1 (top) and Q7 (bottom), to be visually compared with monitoring in \cref{fig:app:postselectioninfig3:monitoring}. With software recovery in panel \cref{fig:app:postselectioninfig3:timeseries}(a), the difference between mitigated and unmitigated expectation values declines to zero as readout error on Q1 increases in \cref{fig:app:postselectioninfig3:monitoring}(b).  For the persistently high readout error for Q7 in \cref{fig:app:postselectioninfig3:monitoring}(e), there is no difference between unmitigated and mitigated values in \cref{fig:app:postselectioninfig3:timeseries}(c).

We speculate that TLS-induced readout error \cite{thorbeck2023readoutinduced} and/or classical state discrimination errors on ancilla qubits could increase probability in applying incorrect ancilla-controlled recovery operations, as explored in rightmost column of \cref{fig:app:postselectioninfig3:timeseries}. The case where all samples are retained (left column) is compared with the case where samples are post-selected on the `00' outcome on both ancilla qubits (right column). Using post-selection appears to enhance the mean performance differential between mitigated (red, blue) and unmitigated (yellow) approaches from the same raw dataset in \cref{fig:app:postselectioninfig3:timeseries}(b),(d).  Since the circuit under consideration produces any of the outcomes $00,01,10,11$ with probability $25$\% in the noiseless case, the effect of post-selection is to discard $\sim 75$\% of the data resulting in larger bootstrapped errors bars in the rightmost column.

We clarify that while post-selection can be a diagnostic tool for probing the performance of measurement-based circuits, post-selection not only reduces the number of eligible mitigation samples but also increases the sampling overhead required by (non-linearly) re-normalizing the post-measurement state. Therefore combining post-selection for use in conjunction with (M)PEC methods in target applications is not intended.

For completeness, we analyze the expectation value over the full database when both Q1 and Q7 are subject to noise, rather than just the first 7 trials where Q1 appears to be performing well. In \cref{fig:app:postselectioninfig3:expvals}(a), the expectation values and standard deviation are computed over temporal trials, where each trial consists of a cycle of learning and mitigation. In this case, error bars reflect the temporal drift in the device. In \cref{fig:app:postselectioninfig3:expvals}(b) expectation values and standard deviations are estimated by bootstrapping the combined database over all cycles of learning and mitigation. Here, the error bars capture shot noise associated with the number of samples used during mitigation. The noise strength for each cycle of learning and mitigation is reported in \cref{fig:app:postselectioninfig3:expvals}(c).

\section{Supporting analysis for Fig. 4 \label{app:supportingdata:fig4}}
We revisit the results of Fig. 4 by examining where explicit recovery operations were not applied by the controller.
\begin{figure}[h]
    \centering
    \includegraphics[scale=0.9]{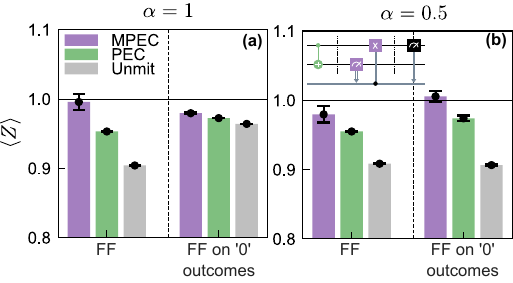}
    \caption{Error mitigated expectation values $\langle Z \rangle$ reported in Fig. 4 for initial states $\alpha=1, 0.5$.  (a),(b) Using explicit feedforward (`FF', left), as reported in Fig.4 (b),(e), one compares to post-selection where samples are retained only if a classically controlled $X$ gate was never triggered (right). Observed accept rate (a) $0.94 \pm 0.15$ where ideal probability of applying a $X$ recovery operation is zero, and (b) $0.50 \pm 0.05$,  where ideal probability for $X$ recovery operation is half.}
    \label{fig:app:postselectioninfig4:expvals}
\end{figure}

In \cref{fig:app:postselectioninfig4:expvals}, results of hardware feedforward in the main text are re-stated and compared to the case when circuit samples are post-selected if no recovery operation is applied. In all cases, circuits incur idling error associated with classical communication with the real-time controller. Panels (a) and (b) differ by the initial state on the data qubit during mitigation, but the choice of initial state does not impact the learning protocol. 

Focusing on the unmitigated but twirled circuits (gray), a performance improvement under post-selection is seen in \cref{fig:app:postselectioninfig4:expvals}(a), where the ideal probability of applying a classically controlled $X$ operation is zero. In \cref{fig:app:postselectioninfig4:expvals} (b) the ideal probability of applying a classically controlled $X$ operation is exactly half, and no such improvement in the performance of the unmitigated but twirled circuits is observed. Noting that the measurement channel is twirled, and readout errors on ancilla measurements are symmetrized, we speculate that classical errors on the feedforward control wire for the unmitigated case may give rise to such a signature and investigating these errors remains the subject of future work.



 \end{document}

%% file: preamble.tex
\usepackage{appendix}
\usepackage{xcolor}
\usepackage{graphicx}
\usepackage{amsmath,amssymb,amsfonts}
\usepackage{booktabs}
\usepackage{multirow}

\usepackage{cleveref}
\crefname{equation}{Eq.}{Eqs.}
\crefname{figure}{Fig.}{Figs.}
\crefname{table}{Table}{Tables} 
\crefname{section}{Section}{Sections}
\crefname{chapter}{Chapter}{Chapters}
\crefname{appendix}{Appendix}{Appendices}
\crefname{algorithm}{Algorithm}{Algorithms}
\crefname{theorem}{Theorem}{Theorems}
\crefname{defn}{Definiton}{Definitions}
\crefname{azm}{Assumption}{Assumptions}
\crefname{corollary}{Corollary}{Corollaries}
\crefname{lemma}{Lemma}{Lemmas}
\crefname{thmprop}{property}{properties}
\crefname{proposition}{Proposition}{Propositions}
\crefname{remark}{Remark}{Remarks}